\documentclass{article}

\usepackage{subfig}
\usepackage{caption}
\usepackage{graphicx}
\usepackage{url}
\usepackage{amsmath}
\usepackage{latexsym}

\newcommand{\fix}[2]{{\bf FIX}\footnote{{\bf #1:} #2 }}
\renewcommand{\fix}[2]{}

\newcounter{example}[section]
\newenvironment{example}[1][]{\refstepcounter{example}\par\medskip
	\noindent \textbf{Example~\theexample. #1} \rmfamily}{\medskip}

\begin{document}

\title{Don't Tell Me The Cybersecurity Moon \\ Is Shining... \\
(Cybersecurity Show And Tell)}
%

\author{Luca Vigan\`o \\[0.1em] 
Department of Informatics \\
King's College London, London, UK \\ 
luca.vigano@kcl.ac.uk}

\date{\today}

\maketitle

\begin{abstract}
``Show, don't tell'' has become the literary commandment for any writer. It applies to all forms of fiction, and to non-fiction, including scientific writing, where it lies at the heart of many scientific communication and storytelling approaches. In this paper, I discuss how ``show \emph{and} tell'' is actually often the best approach when one wants to present, teach or explain complicated ideas such as those underlying notions and results in mathematics and science, and in particular in cybersecurity. I discuss how different kinds of artworks can be used to explain cybersecurity and I illustrate how telling (i.e., explaining notions in a formal, technical way) can be paired with showing through visual storytelling or other forms of storytelling. I also discuss four categories of artworks and the explanations they help provide.
\end{abstract}


\section{Show, don't tell!}

In May 1886, the Russian playwright and short-story writer Anton  Chekhov wrote a letter to his brother Alexander, who too had literary ambitions, providing him with the following advice: 
\begin{quote}
	In descriptions of Nature one must seize on small details, grouping them so that when the reader closes his eyes he gets a picture. For instance, you'll have a moonlit night if you write that on the mill dam a piece of glass from a broken bottle glittered like a bright little star, and that the black shadow of a dog or a wolf rolled past like a ball. \cite{UnknownChekhov}
\end{quote}
This is often misquoted as
\begin{quote}
	Don't tell me the moon is shining; show me the glint of light on broken glass.
\end{quote}
but the misquote is understandable: it is crisper and, essentially, preserves the intended meaning. 
It is a concise injunction that has become the literary commandment for any writer: \emph{show, don't tell}!

The distinction between telling and showing was popularized by the literary scholar Percy Lubbock in his 1921 book ``The Craft of Fiction''~\cite{Lubbock1921}.
\emph{Show, don’t tell} is a writing technique in which story and characters are related to the reader through action, words, dialogues, thoughts, senses, and feelings rather than through the author's exposition and description. In a nutshell: telling states, showing illustrates. 

Several other literary scholars and writers have since then discussed the \emph{show, don't tell} style of writing, including Ernest Hemingway and Stephen King, two of the style's most prominent proponents.
%
For instance, in his memoir ``On Writing: A Memoir of the Craft'', Stephen King writes:
\begin{quote}
	Description is what makes the reader a sensory participant in the story. [...] Description begins with visualization of what it is you want the reader to experience. It ends with your translating what you see in your mind into words on the page. It's far from easy. 
	%
	[...]
	Thin description leaves the reader feeling bewildered and nearsighted. Overdescription buries him or her in details and images. The trick is to find a happy medium. \cite[pp.~173--174]{KingWriting}
\end{quote}

\emph{Show, don't tell} applies also to all forms of fiction (including poetry, scriptwriting and playwriting) and to non-fiction (including speech writing and blogging). 
Does it apply also to scientific writing? Of course it does. In fact, even though I am not aware of any explicit theoretical (or practical, for that matter) investigation of the use of \emph{show, don't tell} in scientific writing, the \emph{show, don't tell} spirit lies at the heart of many successful scientific communication and storytelling approaches, such as those discussed in John Brockman's ``The Third Culture: Beyond the Scientific Revolution''~\cite{Brockman95}\footnote{See also Edge.org, 
	the website of the Edge Foundation, Inc., which was launched in 1996 as the online version of ``The Reality Club''
	to display the activities of ``The Third Culture.''}. Storytelling has been used widely, and very successfully, as a pedagogical device in textbooks and science outreach endeavors, e.g., \cite{Capozucca2018,Erwig2017,Gouthier2019,Harding2018,Mehlmann2000} to name a few. 
Rina Zazkis and Peter Liljedahl, in particular, have been instrumental in promoting storytelling in the mathematics classroom. They begin their book ``Teaching Mathematics as Storytelling'' by writing:
\begin{quote}
	We like to tell stories. We tell stories about mathematics, about mathematicians, and about doing mathematics. We do this firstly because we enjoy it. We do it secondly because the students like it. And we do it thirdly because we believe that it is an effective instructional tool in the teaching of mathematics. We are not alone in this. There is ample literature to support the enjoyment of storytelling on the part of both the story teller and the story listener. There is also an abundance of anecdotal data that suggest “telling a story creates more vivid, powerful and memorable images in a listener’s mind than does any other means of delivery of the same material”~\cite[p.~xvii]{Haven2000}. 
	Aside from the educational value, however, there is also beauty. There is beauty in a story well told, and there is beauty of a story that can move a listener to think, to imagine, and to learn. \cite[p.~ix]{ZazkisLiljedahl2009}
\end{quote}
I find this remark about beauty particularly fascinating, especially since my colleague Giampaolo Bella and I have been reflecting about beauty in security~\cite{BellaVigano2015}, which has led us to work with Karen Renaud and Diego Sempreboni to investigate the beautification of security ceremonies (i.e., protocols that are executed by machines and human users)~\cite{bella2019investigation}.
We are currently working on a deeper analysis of the role that beauty plays in security, but I am digressing from the main topic of this paper, so let me return to Zazkis and Liljedahl, who, a few lines after the quote above, discuss the purpose of telling stories in the classroom:
\begin{quote}
	We tell stories in the mathematics classroom to achieve an environment of imagination, emotion, and thinking. We tell stories in the mathematics classroom to make mathematics more enjoyable and more memorable. We tell stories in the mathematics classroom to engage students in a mathematical activity, to make them think and explore, and to help them understand concepts and ideas. \cite[p.~ix]{ZazkisLiljedahl2009}
\end{quote}
They quote Egan:
\begin{quote}
	Telling a story is a way of establishing meaning. \cite[p.~37]{Egan1986}
\end{quote}
and then talk about the power of images:
\begin{quote}
	One result of the development of language was the discovery that words can be used to evoke images in the minds of their hearers, and that these images can have as powerful emotional effects as reality might, and in some cases even more. \cite[p.~15]{ZazkisLiljedahl2009}
\end{quote}

In fact, when Egan, Haven, Zazkis and Liljedahl, as well as many others, speak of telling, and then, more concretely, of storytelling, they invoke images and imagery. One would be tempted to cite the old adage ``A picture is worth a thousand words.''\footnote{In 1921, the advertising trade journal ``Printer's Ink'' published an article by Frederick R. Barnard titled ``One Look is Worth a Thousand Words'' in which Barnard claims that the phrase has Japanese origin. But in 1927, ``Printer's Ink'' published an advert by Barnard with the phrase ``One Picture Worth Ten Thousand Words,'' where it is labeled a Chinese proverb. The Japanese and Chinese attributions were meant to give it more credibility, a sense of gravitas and a touch of mystery and philosophy, so much so that the proverb is nowadays commonly, and wrongly, attributed to the Chinese philosopher and politician Confucius.} So, if storytelling is powerful, and images and pictures even more so, why not combine them? Why not tell \emph{and} show? Or, better, \emph{show and tell}?

\section{Show and tell}

Dan Roam has written a number of books, including ``The Back of the Napkin"~\cite{Roam2008} and "Show and Tell"~\cite{Roam2014}, in which he has been proposing visual thinking and storytelling for problem solving.\footnote{``Show and tell'' is also the name of a common classroom activity in elementary schools, especially in English-speaking countries, in which a child brings an item from home and explains to the class why he/she chose that item and other relevant information. This activity is useful also for adults~\cite{Nelson-Show}, but it is quite different from the \emph{show and tell} that Roam champions and the one that I discuss here.} Roam is also an engaging speaker and some of his presentations are available online. In particular, when he presented ``The Back of the Napkin'' at Google~\cite{RoamGoogle2008}, he said:
\begin{quote}
	Any problem can be clarified significantly, if not outright solved, through the use of a picture. 
\end{quote}
Drawing on his collaborations with visual scientists and neurobiologists, he added:
\begin{quote}
	If we can take the time to use these simple pictures to	help us figure out what we're talking about, we now have this incredibly powerful tool to use to share with other people when we meet them, and the beauty of it is that they're not going to forget what we told them. 
	[...]
	When we draw in front of someone at the same time that we're talking, 
	[...]
	we are actually activating processing centers in the brain that are really, 
	really excited. Our brain wants to get information visually as well as verbally 
	[...]
	when we draw the picture at the same time that we're talking, they get it and it's like manna for the person's brain. This is the way the brain wants to process information.  
	[...]
	The picture is something that is archival and can be taken along, and we essentially guarantee that the person we gave it to, that we drew this picture for, really does understand what we were talking about [...] almost invariably we can guarantee that they understood it in exactly the way we meant because we created the picture with them. 
\end{quote}

In this talk, and in his books, Roam proposes to draw pictures in real time when presenting an idea, when addressing a problem and pitching its possible solution. This is one of the possible ways in which one can realize 
\emph{show and tell}. In this paper, I will mainly explore another way, namely the use of existing artworks (films, in particular, but not only). The idea is that while \emph{show, don't tell} is the commandment for fiction, in the case of non-fiction, \emph{show and tell} is often the best approach when one wants to present, teach or explain complicated ideas such as those underlying notions and results in mathematics and science. Or in cryptography and cybersecurity, which is my own discipline.

\begin{figure}[t!]
	\begin{center}
		\includegraphics[scale=0.41]{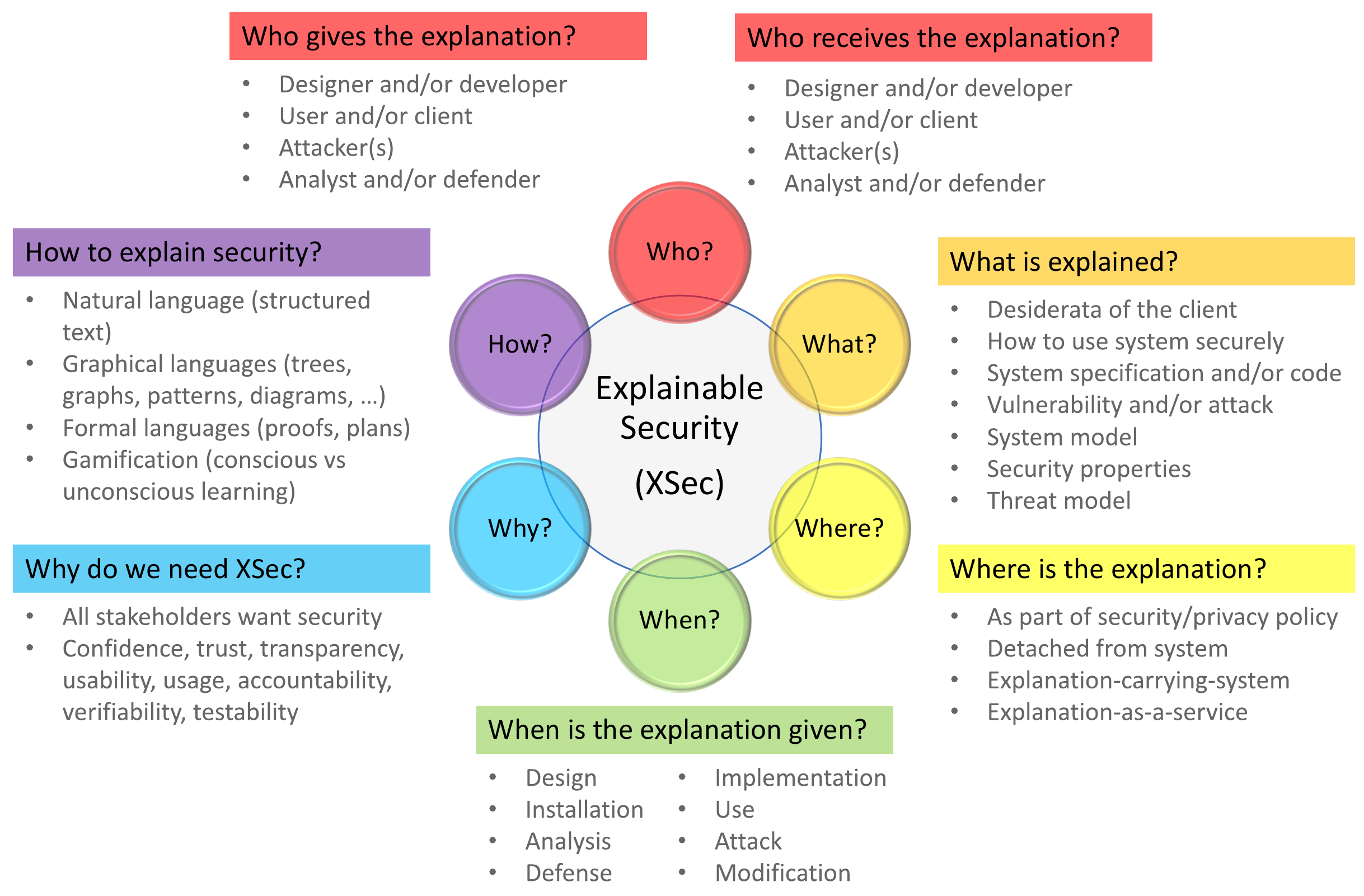}
		\caption{The Six Ws of Explainable Security (from~\cite{XSec2020})}
		\label{fig:XSec}
	\end{center}
\end{figure}

In my paper ``Explaining Cybersecurity with Films and the Arts'' in ``Imagine Math 7''~\cite{ViganoFilm}, I discussed how, in the context of research in \emph{Explainable Security (XSec)}~\cite{XSec2020}, popular movies and other artworks can be used to explain a number of basic and advanced cybersecurity notions, ranging from 
security properties (such as anonymity, pseudonymity and authentication) to the algorithms, protocols and systems that have been developed to achieve such properties, and to the vulnerabilities and attacks that they suffer from.
As this paper is a natural continuation of~\cite{ViganoFilm}, let me repeat some text that I wrote there and then expand on it:
\begin{quote}
	In~\cite{XSec2020}, we discussed the ``Six Ws'' of XSec (Who? What? Where? When? Why? and How?, as summarized in Fig.~\ref{fig:XSec}) and argued that XSec has unique and complex characteristics: XSec involves several different stakeholders (i.e., system's developers, analysts, users and attackers) and is multi-faceted by nature, as it requires reasoning about system model, threat model and properties of security, privacy and trust as well as concrete attacks, vulnerabilities and countermeasures. 
\end{quote}
This paper, like~\cite{ViganoFilm}, is mainly about the ``Who?'' and the ``How?''.
As pointed out in~\cite{XSec2020}, the recipients of the explanations might be so varied, ranging from experts to laypersons, that they require quite radically different explanations, formulated using different languages. 
Experts typically only accept detailed technical explanations, whereas laypersons are often scared off by explanations of cybersecurity (say, how to interact with a system or an app) that are detailed but too technical. Such an explanation might even repulse the laypersons and make them lose all trust in the explanation and, ultimately, in the cybersecurity of the system that is being explained. This repulsion and lack of trust might lead to users interacting with systems in ways that, unbeknownst to the users and possibly even to the developers and administrators of the systems, are vulnerable to attacks (to the systems and to the users themselves).
In practice, however, laypersons are rarely given explanations that are tailored to their needs and their ability to understand.

As discussed
in~\cite{ViganoFilm}, clear and simple explanations with popular films and the arts allow experts to target the laypersons, reducing the mental and temporal effort required of them and increasing their understanding, and ultimately their willingness to engage with cybersecurity systems. In other words, with reference to the research roadmap for XSec that was laid out in~\cite{XSec2020}, this paper and~\cite{ViganoFilm} focus on 
\begin{itemize}
	\item \textbf{Who?} The experts (the system developers but possibly also independent third parties) provide explanations to the laypersons.
	\item \textbf{How?} Using popular films and the arts. (It would also be interesting to consider ``inventing'' new artworks, possibly in real time, like Roam's drawings, and I will return to this in the concluding remarks.)
\end{itemize}
The other ``Ws'' should, of course, be considered too:
\begin{itemize}
	\item \textbf{What?} Everything that pertains to cybersecurity, such as properties, algorithms and protocols, threats, vulnerabilities and attacks. Paraphrasing Roam's assertion quoted above,
	the problem that we want to tackle, and in some cases hopefully outright solve, is the clarification of cybersecurity notions through the use of static or moving pictures (i.e., films) and other artworks.
	\item \textbf{Where?} These explanations could actually be anywhere, either co-located with the system or detached from it. 
	\item \textbf{When?} As these explanations mainly target the system users, the explanations will be given when (or after) the system is deployed, but it will also be useful for the system developers to start working on the explanations at system design time (unless the explanation is provided by a third party, independent of the developers). Such explanations are also an effective instructional tool in the teaching of mathematics, as advocated by Zazkis and Liljedahl~\cite{ZazkisLiljedahl2009} and as I have been doing for  20 odd years, using films and artworks during lectures and in public engagement talks on cybersecurity.
	\item \textbf{Why?} To increase the laypersons' understanding, confidence, trust (and more) in cybersecurity.
\end{itemize}
There is also a second ``Why?'', namely why use popular films and artworks to explain cybersecurity?
Because the added power of telling (i.e., explaining notions in a technical way) \emph{and} showing (via visual storytelling or other forms of storytelling) can help experts to convey the intuition in addition to the technical definition. 

I have given some examples in~\cite{ViganoFilm} considering security properties such as authentication, anonymity, unobservability and untraceability and some algorithms and mechanisms to achieve them (or to attack them). Let me give here another example about anonymous communication, or, more specifically,  unobservability of message exchanges and untraceability of messages.

\begin{example}
	Assume that Alice wants to send a message to Bob but she 
	does not want the attacker Charlie to trace the communication, i.e., Alice (and possibly Bob too) does not want Charlie to observe that Alice sends the message and that the message is received by Bob. The point here is not to keep the contents of the message confidential (that can be achieved by encryption, if needed) but rather that the communication Alice---Bob is confidential, i.e., that Charlie is not able to trace the message and observe that the communication is taking place. 
	There might be plenty of reasons to keep a communication confidential. For instance, Alice might be an employee of Charlie's company and would not want Charlie to know that she has applied for a job at Bob's company, or Charlie might be a crime lord, Alice a snitch and Bob the Police. In such cases, and many others, it is therefore in Charlie's interest to monitor the network over which messages are sent and received (say it is the Internet) and trace messages as they move from one machine to the other in the network, from sender to receiver. And it is in Alice's interest (and possibly Bob's too) to find a security solution to impede Charlie to carry out such tracing, even when it is assumed that Charlie is able to monitor the whole network, as is typically assumed in security analysis, where one considers the most powerful attacker possible, e.g., one who is able to monitor the whole Internet. If the adopted security solution 
	is able to withstand the attacks of such most powerful attacker, then it will for sure be able to withstand also the attacks of a real, and thus less powerful, attacker.
	
	Some security solutions for anonymous communication over a network (and unobservability and untraceability of messages) have been implemented, such as \emph{Mix Networks}~\cite{Chaum81} and \emph{Onion Routing}~\cite{Onion}, but 
	they 
	are among the security solutions that are most difficult to explain from a technical point of view. Similarly, one can give a natural-language definition of untraceability, 
	\begin{quote}
		Untraceability of an object during a process under observation of an attacker is the property that the attacker cannot follow the trace of the object as it moves from one participant or location to another.~\cite{untraceability}
	\end{quote}
	and then refine this initial definition by providing a more formal one, a mathematical one, but I am fairly certain that some readers have already found the natural-language definition to be quite difficult to digest, and they would find the technical definition and the solutions even more impenetrable.
	
	The intuition behind the property and the solution is, however, quite simple. So, rather than \emph{telling} you that the Mix-Network moon is shining, let me first \emph{show} you the glint of light on untraceable communications. Once you get the intuition, the technical explanation will hopefully be much more understandable (if you are, say, a cybersecurity student learning how Mix Networks work) or maybe not needed at all (if you are a layperson interested in understanding why you should trust Mix Networks but not so much in their inner workings). In fact, I will not tell you the technical explanation in this paper, but only show how and why it works.

	Consider the network delimited by the dotted line in Fig.~\ref{fig:mix0}, where the squares represent machines that distribute messages in the network,
	and meet Alice \includegraphics[scale=0.2]{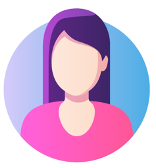} and Bob \includegraphics[scale=0.2]{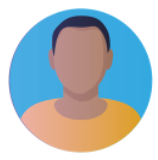}. I don't show the attacker Charlie explicitly, but you can picture him at bird's-eye view, observing the whole network. If Charlie is able to ensure Alice's message is the only one in the network, as in Fig.~\ref{fig:mix1}, then tracing the communication is a trivial task. It is as if the police were chasing a car on a highway, and that car is the only one on that highway.
	
	In fact, to \emph{show} this, consider the photo in Fig.~\ref{fig:OJS}. On June 17, 1994, former NFL player O.J.~Simpson was formally charged with the murders of his ex-wife, Nicole Brown Simpson, and her friend Ron Goldman. Instead of turning himself in, Simpson drove off into a white 1993 Ford Bronco SUV and became a fugitive of the law. The low-speed car-chase that ensued was watched live by an estimated 95 million people and indeed the police closed the highways so that they could easily follow the car for about 60 miles until Simpson's home.
	
	The car-chase would have been more challenging had there been other cars on the highway (instead of having been forced to pull over as shown in the photo). So, let's add some more agents who send and receive messages alongside Alice and Bob  (the machines in the network are also allowed to send messages),
	as shown in Fig.~\ref{fig:mix2}. Charlie's task is now more complex, but still feasible: if he wishes to find out who Alice is communicating with, Charlie just needs to follow the messages that are sent by Alice to the first machines in the network, and then follow the messages that are sent by these machines, and so on, until he has identified all possible traces from Alice to the possible recipients. Basically, this means dispatching one police car for every suspect car. A powerful attacker would certainly have enough police cars at his disposal. Hence, to achieve untraceability, we need suspect cars, lots of suspect cars. That is, we need many more agents sending and receiving many more messages as shown in Fig.~\ref{fig:mix3}. Charlie would now need to follow all of the messages, and the more there are, the harder Charlie's task as Alice's message could be any one of the messages that are circulating in the network. In technical terms, this set of messages is called the \emph{anonymity set}: Alice's communication with Bob is anonymous as Alice's message is not identifiable within the set of messages. Of course, the larger the set, the higher the level of anonymity.

	\begin{figure}[t!]
		\begin{center}
			\includegraphics[scale=0.225]{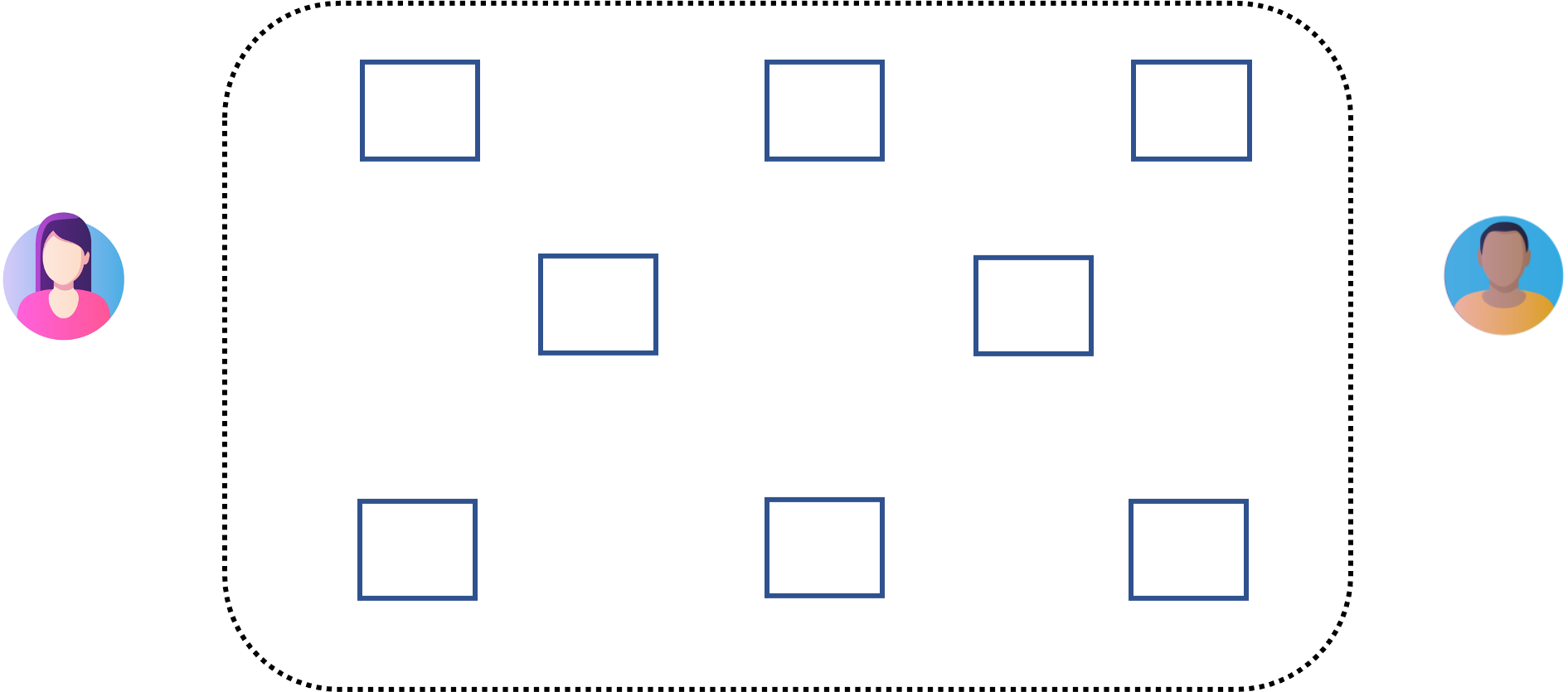}
			\caption{Alice, Bob and the network}
			\label{fig:mix0}
		\end{center}
	\end{figure}
	
	\begin{figure}[t!]
		\begin{center}
			\includegraphics[scale=0.225]{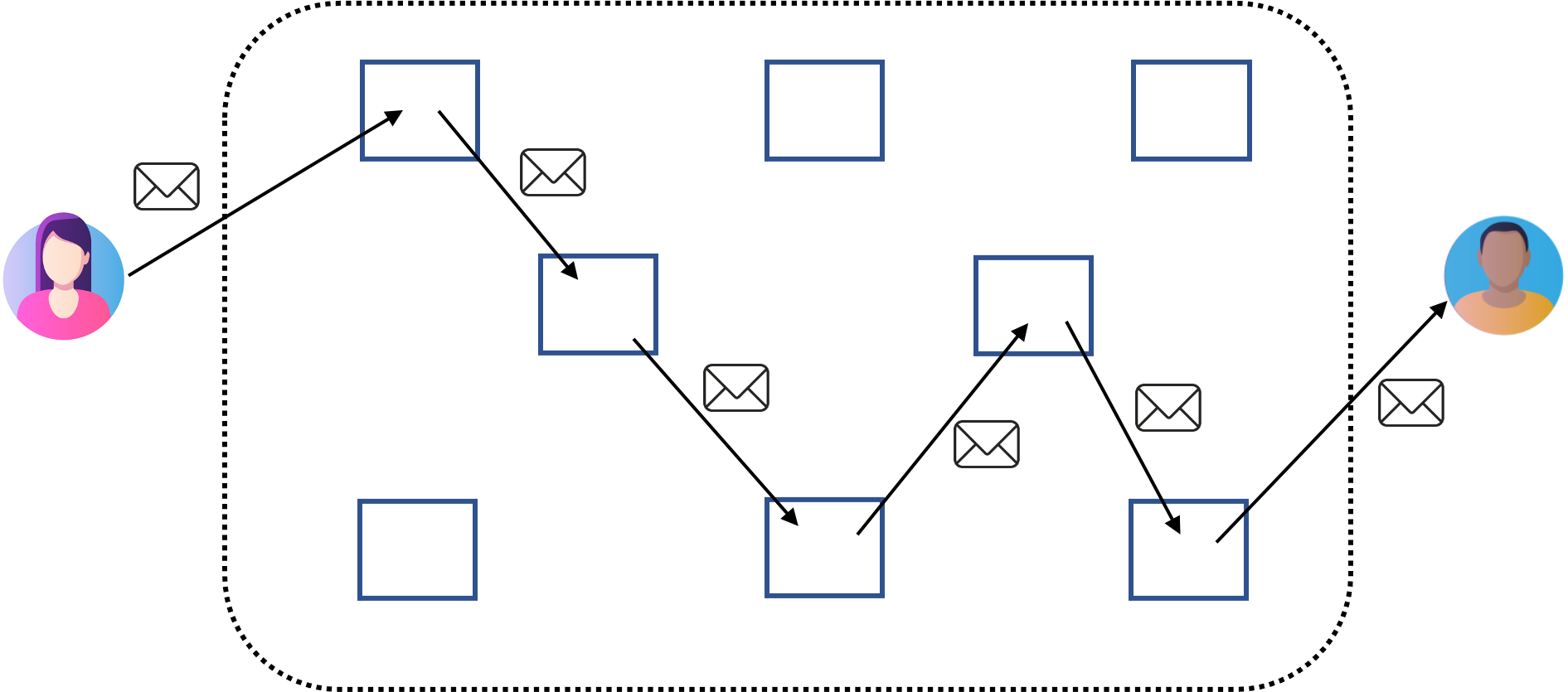}
			\caption{Alice sends a message to Bob... but it is the only message in the network and thus can be easily traced}
			\label{fig:mix1}
		\end{center}
	\end{figure}
	
	\begin{figure}[t!]
		\begin{center}
			\includegraphics[scale=0.225]{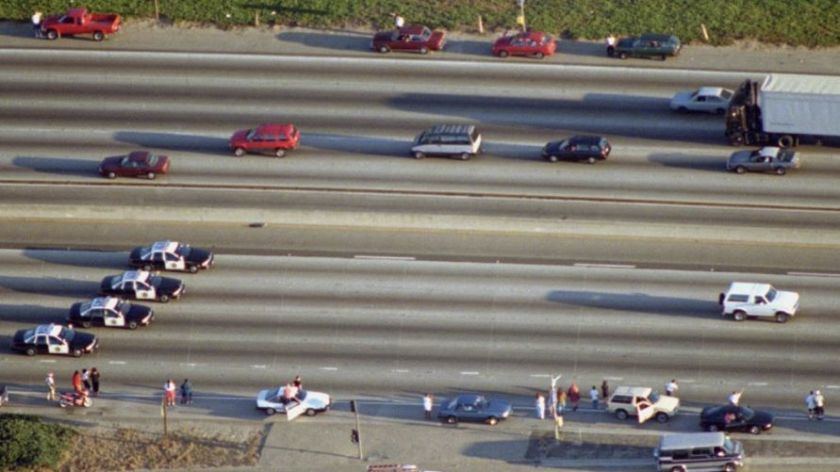}
			\caption{The police chasing O.J.~Simpson's white Ford Bronco (It was impossible to trace the author of this photo even though many newspapers online have used it.)}
			\label{fig:OJS}
		\end{center}
	\end{figure}

	\begin{figure}[t]
		\begin{center}
			\includegraphics[scale=0.225]{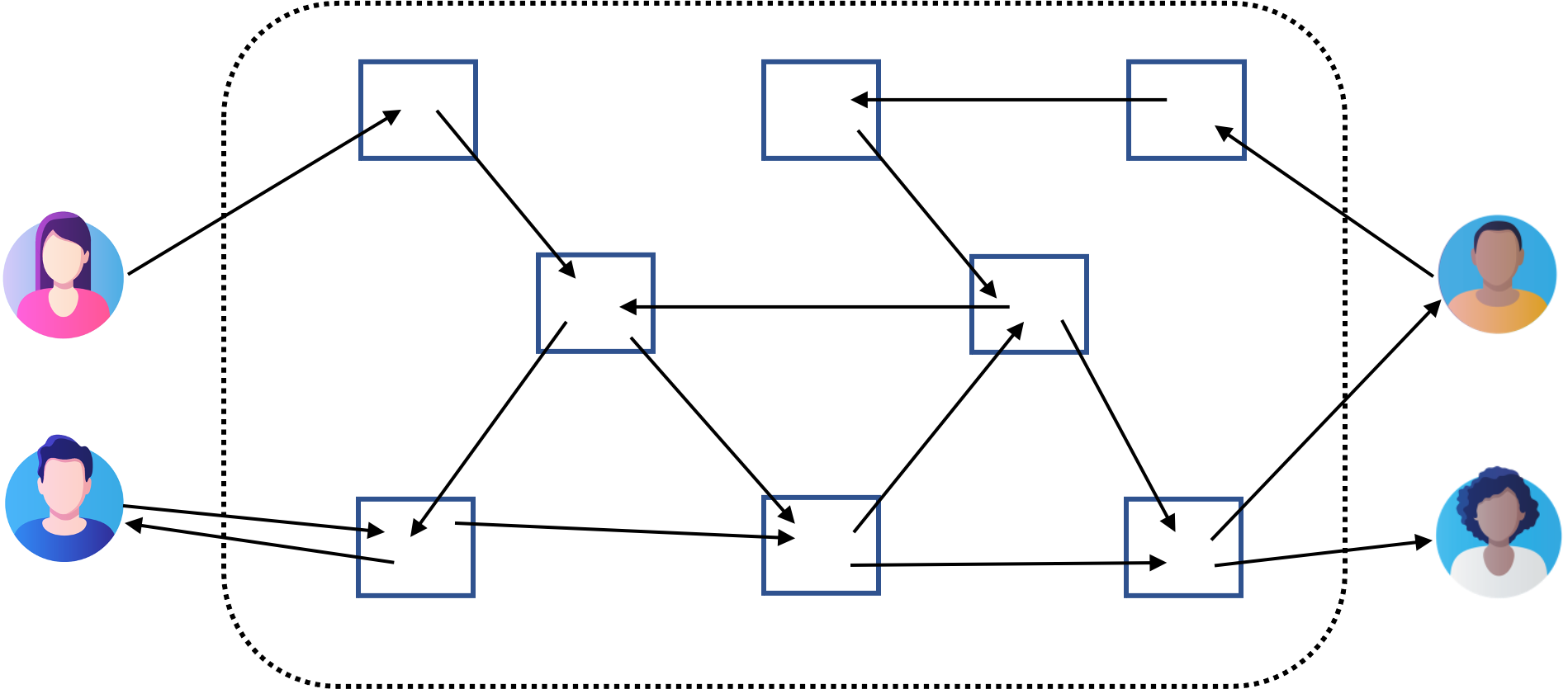}
			\caption{Some more agents, some more messages}
			\label{fig:mix2}
		\end{center}
	\end{figure}
	
	\begin{figure}[t]
		\begin{center}
			\includegraphics[scale=0.225]{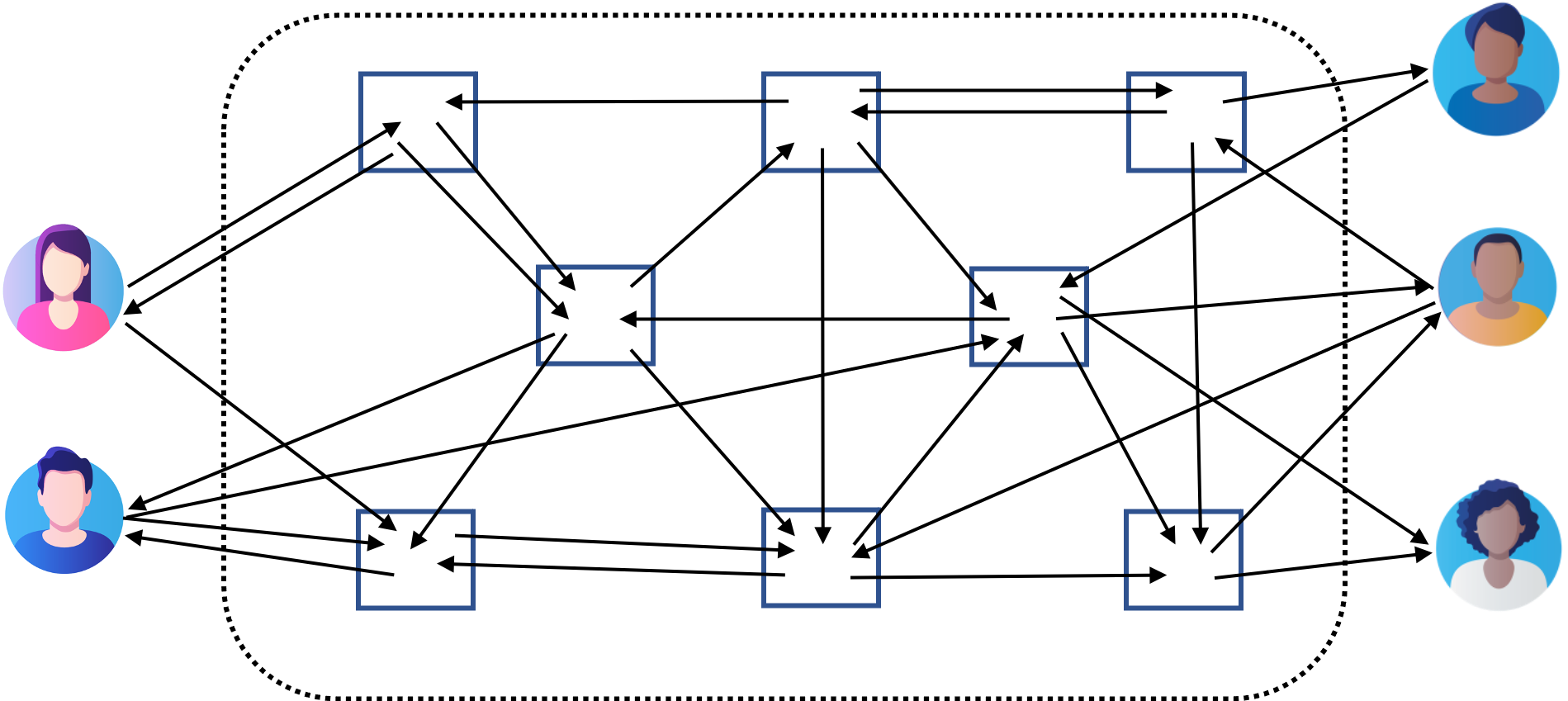}
			\caption{Towards untraceability}
			\label{fig:mix3}
		\end{center}
	\end{figure}
	
	I have already discussed anonymity sets in~\cite{ViganoFilm} using the ``I'm Spartacus!'' scene in Stanley Kubrick's ``Spartacus''~\cite{Spartacus} and the climatic museum scene in ``The Thomas Crown Affair''~\cite{ThomasCrown1999} (and more).
	As I observed in~\cite{ViganoFilm}, the main idea underlying the way in which Mix Networks realize untraceability of messages is that there are plenty of messages circulating in a network and that such messages all look alike (in the sense that they are all plausible messages). But Mix Networks do more than that. The machines in the network are called ``Mixes'' since they receive a batch of messages in input, mix them and then output them in different order, so that an attacker who is observing the input and output of each Mix, but cannot open the Mixes to look at the inner working, cannot associate output messages with input messages. There is no first-in/first-out or last-in/last-out association (nor any other association such as first-in/last-out). The first message that is output by a Mix could correspond to any of the messages that the Mix received in input. Moreover, Mixes also ensure that there is always enough traffic in the network by sending ``dummy messages'' (i.e., fake messages that are then discarded) and they require that all messages have the same size.
	
	How can we ``show'' that? For instance, by using a scene from the movie ``Baby Driver''~\cite{BabyDriver}. The movie begins with 
	a bank heist and the ensuing car chase: three bank robbers are escaping in a red sedan car (a 2006 Subaru WRX STI in San Remo Red) chased by several police cars and a helicopter. Baby, the robber's getaway driver, has a brilliant idea: he enters a highway by going the wrong way into oncoming traffic\footnote{This is an homage to another great car chase, the one in ``To Live and Die in L.A.''~\cite{ToLiveandDie}.} and then performs a u-turn to merge into normal traffic when he spots two other similar red sedan cars proceeding side by side. Aerial shot from the helicopter: three red cars side by side, with Baby's car in the middle. But when the cars enter a short tunnel, Baby performs a mix: he drives in front of the car on his left and breaks suddenly, forcing that car to take Baby's place in the middle of three, so that, when they exit the tunnel, the helicopter continues to chase the car in the middle and Baby's car can safely escape by taking the next exit.
	Traffic, an anonymity set consisting of similar, if not identical, cars and a mix changing the order of the ``messages'': Baby gets away using a small Mix Network. 
	\hfill $\Box$
\end{example}

This lengthy example has hopefully demonstrated that even challenging concepts such as untraceability and Mix Networks can benefit from the \emph{show and tell} treatment. I urge you to watch the movie scenes that I have referenced above (and the ones I will mention below) as doing so will bear witness to the fact that the synergy between showing and telling adds another dimension to what is told, a sensory one based on visual storytelling.\footnote{This is reminiscent 
	of the way in which a musical score adds an emotional layer to the images of a film, thus contributing in a fundamental way to the storytelling. This has been explained brilliantly by Stewart Copeland in the second episode, aptly titled ``Telling Tales'', 
	of the documentary~\cite{Copeland}, 
	in which 
	Copeland discusses music in films with composer Danny Elfman:
	\begin{description}
		\item[Copeland:] Why do the directors need this? They're telling a perfectly good story, with a perfectly terrifying antagonist, a handsome protagonist, a beautiful love interest. Why do they need music?
		\item[Elfman:] Because music does something they learned very early on, that the pictures couldn't do. 
		\item[Copeland:] Take the decidedly lukewarm chills of early horror movies, for example.
		\item[Elfman:] In the very first Frankenstein and the first Dracula, no music. All music was, in the first films, was opening and closing, like a play, and then they figured out a few years later, 1933 and 1935, ``Why don't we take it up a level?'' If you put this dramatic music, it really raises the stakes.
		\item[Copeland:] As shown in in the pioneering movie King Kong. 
		\item[Elfman:] And if you put something heartbreaking when, you know, your hero or heroine is going to die, it really raises the stakes. [...] It goes straight to the heart. 
	\end{description}
	In addition to ``Frankenstein''~\cite{Frankenstein}, ``Dracula''~\cite{Dracula} and ``King Kong''~\cite{KingKong}, Copeland and Elfman then also discuss on how Bernard Herrmann's score punctuates and amplifies Alfred Hitchcock's images in the movie ``Vertigo''.}




As discussed in~\cite{ViganoFilm} and above, a clear and simple explanation, with something that they are already familiar with, such as a non-security related movie or novel, will help make laypersons less irritated, stressed and annoyed, and thus more receptive. 
Different from~\cite{ViganoFilm}, in the remainder of this paper I take a more systematic approach, mapping different approaches to cybersecurity \emph{show and tell}. 

Zazkis and Liljedahl consider stories that frame or provide the background for a mathematical activity, and they distinguish between stories that introduce, and stories that accompany and intertwine with mathematical activity. 
In particular, they discuss the following categories of stories for the teaching of mathematics:
\begin{itemize}
	\item \emph{Stories that set a frame or a background}, i.e., stories in which hero(in)es have to overcome obstacles to reach their goal (e.g., Oedipus solving the riddle of the Sphinx), stories of secret codes (e.g., stories in which decoding a message can save lives, or point to a treasure, win a princess’ heart, or ensure fame and glory), and stories of treaties or contracts (e.g., the ``contract'' that Multiplication and Division shall be performed before Addition and Subtraction, but in the order in which they appear in any calculation).
	\item \emph{Stories that accompany the content} (e.g., derived from the history of mathematics and the lives of mathematicians) and \emph{stories that intertwine with the content} in which mathematical content emerges through the story, at times leaving the story behind and at times staying with the story. Zazkis and Liljedahl point out that the latter are harder to find, but still they are able to provide some interesting examples (e.g., using true and apocryphal stories about the Syracusan mathematician Archimedes).\footnote{I have also some experience with this: in the early Noughties I wrote a play about the French mathematician \'Evariste Galois, who was killed in a duel at age 20 in 1832~\cite{Vigano-Galois,Vigano-Galois-eroe05,Vigano-Galois-hero07}. The Teatro Stabile di Genova, which produced the play, had the brilliant idea to schedule morning performances for middle and high school students, and I have been told by many of them that they had never thought that mathematics could be thrilling and moving.}
	\item \emph{Stories that introduce}, i.e., stories that serve well to introduce concepts, ideas or a mathematical activity (e.g., introducing exponential growth through the classical story of grains of rice and the chessboard).
	\item \emph{Stories that explain}, e.g., riddles such as the ``missing dollar'' or `` If a hen-and-a-half lays an egg-and-a-half in a day-and-a-half, how many days does it take one hen to lay one egg?'' Zaskis and Liljedahl support this kind of stories by writing 
	\begin{quote}
		Mathematics is often perceived by learners as a collection of facts and skills; facts and skills that are sometimes seen as counterintuitive. When this happens a common reaction is to seek refuge in the meaningless memorization of rules. Experienced teachers can easily point to such places, places in which encounters with mathematics are most puzzling and rules are most prevalent. Instead of reciting rules, however, we suggest explaining these rules with stories. \cite[p.~51]{ZazkisLiljedahl2009}
	\end{quote}
	which reinforces the point that I tried to make above when discussing the benefits of explaining cybersecurity through artworks, and to which I will return below.
	\item \emph{Stories that ask a question} and encourage the students to engage with the story to arrive at the answer.
	\item \emph{Stories that tell a joke}, since humor can enhance both the telling and the hearing of a story, and thereby indirectly influence learning. Two notable examples about arithmetic and logic are ``Mathematics is made of 50 percent formulas, 50 percent proofs, and 50 percent imagination.'' and ``Hofstadter's Law:
	It always takes longer than you expect, even when you take into account Hofstadter’s Law.''~\cite[p.~152]{Hofstadter}
\end{itemize}
Zazkis and Liljedahl then also discuss how teachers can create a story and they provide a ``planning framework'' demonstrating how instruction of specific mathematical topics or concepts can be planned and implemented.

\section{Cybersecurity show and tell}
So, what about cybersecurity? Without limiting the discussion to teaching, but considering all kinds of learning experiences, including scientific communications and public engagement or outreach activities, we can first of all divide the use of artworks to explain cybersecurity into two broad categories:
\begin{itemize}
	\item \emph{Using existing artworks}.
	\item \emph{Using new artworks that have been created on purpose}.
\end{itemize}
My work so far has focused on the first category, but together with colleagues in computer science and psychology and with artists and curators we have begun tackling also the second category, so let me focus on the first and conclude by providing a few more details about the second.
The existing artworks can be divided into the following 4 sub-categories, for which I provide lists of examples that are by no means exhaustive:
\begin{itemize}
	\item \emph{Artworks about hackers, codebreakers and cybersecurity experts}.
	\item \emph{Artworks about detectives or spies who use or are confronted with cybersecurity problems and solutions}.
	\item \emph{Artworks about ordinary people confronted with cybersecurity or artworks with references to cybersecurity}.
	\item \emph{Artworks that are not explicitly about cybersecurity but can be used to explain cybersecurity notions}.
\end{itemize}

\subsection{``Yes, I am a criminal. My crime is that of curiosity. I am a hacker, and this is my manifesto.''~\cite{Hackers}}
This category includes 
\emph{artworks about hackers, codebreakers and cybersecurity experts}, who make for very interesting hero(ine) or anti-hero(in)es.
In May 2021, a search on the Internet Movie Database (\url{www.imdb.com}) with the keyword ``hacker'' returned 558 titles of films and TV series, many of which were adapted from novels. Notable examples range from faithful or apocryphal biographies of famous codebreakers and cryptologists like Alan Turing~\cite{ImitationGame,Enigma} and John Nash~\cite{BeautifulMind} or less famous or even made-up ones~\cite{SecretsofScotlandYard,Sebastian}, 
to stories of bad ``black-hat'' hackers who carry out cyberattacks or of good ``white-hat'' hackers who save the day such as in the film ``Hackers''~\cite{Hackers} quoted in the title of this category\footnote{That quote was inspired by the article ``The Conscience of a Hacker'' written by the real-life hacker ``The Mentor'' shortly after his arrest~\cite{TheMentor}. The article ends with the following words:
\begin{quote}
	This is our world now... the world of the electron and the switch, the	beauty of the baud.  We make use of a service already existing without paying for what could be dirt-cheap if it wasn't run by profiteering gluttons, and you call us criminals. We explore... and you call us criminals. We seek after knowledge... and you call us criminals.  We exist without skin color, without nationality, without religious bias... and you call us criminals. You build atomic bombs, you wage wars, you murder, cheat, and lie to us and try to make us believe it's for our own good, yet we're the criminals.
		
	Yes, I am a criminal.  My crime is that of curiosity.  My crime is
	that of judging people by what they say and think, not what they look like. My crime is that of outsmarting you, something that you will never forgive me for.
		
	I am a hacker, and this is my manifesto.  You may stop this individual, but you can't stop us all... after all, we're all alike.
\end{quote}	
}
as well as in~\cite{Sneakers,Wargames,Swordfish,TheNet,Fortress,LiveFreeorDieHard,MrRobot,BletchleyCircle,BletchleyCircleSF,CSICyber,BlackHat,Snowden}.
Some are quite realistic in their depiction of cybersecurity, few are real, most are inventive, giving black/white-hat hackers and codebreakers almost superhero-like abilities, but all of them make for a good \emph{show} companion, as a speaker can discuss how faithfully or not cybersecurity has been portrayed. In fact, bad portrayals can be particularly useful to discuss misconceptions and correct possible  prejudices. 

In addition to films and TV series, there are also a few plays about cybersecurity, such as ``Teh Internet Is Serious Business'' (sic!), which provides a fictional account of the hacktivism of the collectives Anonymous and LulzSec and in which coding is ingeniously and amusingly symbolized by means of interpretive dance~\cite{TehInternet}, as well as ``Hackers''~\cite{HackersPlay} and ``The Nether''~\cite{TheNether}.
There are also many novels, such as, to name just the works of some of the most influential authors, the ``Sprawl trilogy''~\cite{Sprawl} by William Gibson, the cyberpunk writer who coined the word ``cyberspace'' in the short story ``Burning Chrome'', Neal Stephenson's ``Snow Crash''~\cite{SnowCrash} and ``Cryptonomicon''~\cite{Cryptonomicon}, Dan Brown's ``Digital Fortress''~\cite{DigitalFortress}, as well as Stieg Larson's ``Millennium'' trilogy~\cite{Millennium} (and the film adaptations that have been filmed in Sweden and the USA~\cite{Girl-original,Girl2-original,Girl3-original,Girl-remake}), and the sequel by David Lagercrantz~\cite{GirlSpider} (which too has been filmed~\cite{Girl4}).
\emph{Showing} with plays and novels might be less immediate and thus more challenging, but that does not mean that it will be less effective.


\subsection{``I have said enough to convince you that ciphers of this nature are readily soluble.''~\cite{TheGoldBug}}
This category includes 
\emph{artworks about detectives or spies who use or are confronted with cybersecurity problems and solutions}. This category is closely connected to the previous one and in many cases there are some overlaps, e.g., serial killers writing in code such as the ``Zodiac Killer''~\cite{Zodiac}, criminal masterminds using steganography to hide their messages~\cite{ConAir}, or the mathematics prodigy who helps his FBI-agent brother solve crimes in the TV series ``NUM3ERS''~\cite{Numb3rs}. 

As pointed out by John F.~Dooley, codes and ciphers have a long history in fiction. In addition to his publications~\cite{Dooley2005,Dooley2016}, Dooley has been collecting a list of pieces of fiction (short stories, novels, chapters in novels, etc.) that contain cryptographic riddles of one sort or another as part of the story line.\footnote{See also the collection of stories of code and ciphers edited by Raymond T.~Bond~\cite{Bond1946}.} 
In fact, several fictional detectives who have had to solve cryptographic riddles, starting with the short story ``The Gold Bug'' by Edgar Allan Poe~\cite{TheGoldBug}, who is generally considered the inventor of the detective fiction genre, but who also had a keen interest in cryptography and in 1841 wrote an essay about secret writing~\cite{PoeSecret}. The most famous private detective of all, Arthur Conan Doyle's Sherlock Holmes, also had to solve several cryptographic riddles, in particular in ``The Adventure of the Dancing Men'' and ``The Valley of Fear''~\cite{TheAdventureoftheDancingMen,TheValleyofFear}, and so had his doubly-fictional sister Enola in Nancy Springer's ``The Enola Holmes Mysteries'' series~\cite{EnolaHolmes} (all of these have also received film and TV adaptations, e.g., \cite{SherlockHolmesNecklace,EnolaHolmesFilm,TheBlindBanker}).
Dilettante detectives have also been tackling cryptographic riddles, e.g., Lord Peter Wimsey in~\cite{HaveHisCarcase}, 
the historian and amateur cryptologist Benjamin Franklin Gates~\cite{NationalTreasure,NationalTreasureBook}, and Robert Langdon, the Harvard professor of history of art and religious symbology\footnote{There is no such things as a professor of symbology in real life, but it is tightly connected to the actual discipline of semiotics, which in turn has been investigated also in the context of cryptography~\cite{Marrone2010}.} created by Dan Brown~\cite{Langdon,TheDaVinciCode,AngelsDemons,Inferno}.

Also these professional and dilettante detectives are typically given abilities that border on the fantastic as they 
``see'' the solutions of the cryptographic riddles in a matter of minutes, if not seconds. They picture them in their head before anyone can and often when nobody else does. This is ``showing'' in a way that is exclusive rather than inclusive.  Similar to the characters of the first category, this facility for cryptography and cybersecurity creates a clear distinction from ordinary people and turns them into true hero(in)es and anti-hero(in)es. Portrayals of spies and secret agents (such as Jason Bourne, the former CIA agent who is the hero of a series of novels, written by Robert Ludlum~\cite{Bourne} and then inherited by Eric Van Lustbader and then by Brian Freeman, and of the subsequent film adaptations directed by Doug Liman, Paul Greengrass and Tony Gilroy) are instead sometimes slightly more realistic, at least in terms of their cryptographic and cybersecurity abilities although maybe not so much in terms of their physical powers, but still almost inevitably go beyond portraying them as simply experts in their field. This is, however, quite understandable, as the story would then otherwise risk being quite dull. Still, using such characters and their stories for \emph{showing} can be very effective also thanks to their worldwide popularity.


\subsection{``I'm in a Secret Club.''~\cite{Peppa}}
This category includes 
\emph{artworks about ordinary people confronted with cybersecurity and artworks with references to cybersecurity}. There are less examples in this category than the previous two, but still several interesting ones, such as: two U.S.~Marines in World War II assigned to protect Navajo Marines as their native language was used as an unbreakable radio code~\cite{Windtalkers}, a businessman whose identity is stolen after he gets a nice call confirming his name and other identifying information~\cite{IdentityThief}, an FBI agent undergoing facial transplant surgery to assume the identity of the criminal mastermind who murdered his only son~\cite{FaceOff}\footnote{``Con Air'', ``National Treasure'', ``Windtalkers'', ``Face/Off''... Nicolas Cage has starred in so many cybersecurity-related movies that he would deserve a dedicated paper, perhaps titled ``Explaining Cybersecurity with Nicolas Cage'' or even better ``Nicolas Cage is the Center of the Cybersecurity Universe.''}, a security specialist forced into robbing a bank to protect his family~\cite{Firewall}, a computer geek inadvertently downloads critical government secrets into his brain and is then recruited by CIA and NSA to help thwart assassins and international terrorists~\cite{Chuck} (the latter two partially belong also in the first category). There are also several animated movies in which cybersecurity features prominently, which suggests that kids, even small children such as those watching ``Peppa Pig'', are probably much less scared by it than adults; for example: multi-factor authentication through biometrics\footnote{\label{footnote-MFA}\emph{Multi-factor authentication (MFA)} is an authentication solution that aims to augment the security of the basic username-password authentication by exploiting two or more authentication factors. In~\cite{ecbstrong}, MFA is for instance defined as:
	\begin{quote}
		a procedure based on the use of two or more of the following elements --- categorised as knowledge, ownership and inherence: i) something only the user knows, e.g., static password, code, personal identification number; ii) something only the user possesses, e.g., token, smart card, mobile phone; iii) something the user is, e.g. a biometric characteristic, such as a fingerprint. In addition, the elements selected must be mutually independent [...] at least one of the elements should be non-reusable and non-replicable.
	\end{quote}	
	The underlying idea is that the more factors are used during the authentication process, the more confidence a service has that the user is correctly identified.} is used both in ``Incredibles 2''~\cite{Incredibles2} and ``Monsters vs.~Aliens''~\cite{Monstersvsaliens}, a guessing/dictionary attack against a password is attempted by Winnie the Pooh's friend Tigger in~\cite{Tigger} by singing the ``Password Song'' with lyrics consisting of every word that Tigger knows, Peppa Pig and their friends use one-time passwords in an episode titled ``The Secret Club''~\cite{Peppa}.

Codes and ciphers (such as the hash function SHA-256, cryptocurrencies and several different ciphers) feature also in animated series for more adult viewers, in particular in Matt Groening's ``The Simpsons''~\cite{TheSimpsons} and ``Futurama''~\cite{Futurama}; many of these examples are discussed by Simon Singh in his book~\cite{Singh-Simpsons} in which he unravels the mathematical secrets of these two series up until 2013.

But it is not just films, TV series and novels. Although examples are more difficult to find, there are also some songs that relate to cybersecurity, such as the song ``Secret'' by The Pierces~\cite{Secret-Pierces}, which references Benjamin Franklyn's famous saying ``Three may keep a secret if two of them are dead'' (which, I believe, in turn references the line ``Two may keep counsel when the third's away'', which is uttered by the villainous Aaron before murdering a nurse to preserve his secret, in William Shakespeare's ``Titus Andronicus'', Act IV, Scene 2):
\begin{quote}
	Got a secret \\
	Can you keep it? \\
	Swear, this one you'll save \\
	Better lock it in your pocket \\
	Takin' this one to the grave \\
	If I show you, then I know you \\
	Won't tell what I said \\
	'Cause two can keep a secret \\
	If one of them is dead ...
\end{quote}  
This category requires substantial work of the presenter to make the connections between \emph{show} and \emph{tell} but, trust me, it can be lots of fun, especially if one shows excerpts of the animated films.

\subsection{``I'm Spartacus!''~\cite{Spartacus}}
This category includes 
\emph{artworks that are not explicitly about cybersecurity but can be used to explain cybersecurity notions}.
This is, in my view, the most interesting category. I have already mentioned above how the ``Baby Driver''~\cite{BabyDriver}, ``Spartacus''~\cite{Spartacus} and ``The Thomas Crown Affair''~\cite{ThomasCrown1999} as well as Alan Moore's graphic novel ``V for Vendetta''~\cite{VforVendetta,VforVendettafilm} can be used for explaining anonymity and untraceability. Other popular films and artworks can be used as metaphors for cybersecurity and in~\cite{ViganoFilm} I discussed how a man-in-the-middle attack can be explained by analogy with the \emph{carte blanche} issued to Milady De Winter by Cardinal Richelieu in ``The Three Musketeers'' by Alexandre Dumas p\`ere~\cite{Dumas}. 

Let me give here some more examples by considering authentication and multi-factor authentication (cf.~Footnote~\ref{footnote-MFA}). Questions of authentication and identity occur extensively in the mythology and fairy-tale literature of most cultures, from Greece to India to China. 
In some cases, authentication occurs by means of passwords, such as the ``Open Sesame'' that opens the mouth of a cave in which Ali Baba and the forty thieves have hidden their treasure in the story ``Ali Baba and the Forty Thieves,'' which first appeared in Antoine Galland's version of ``One Thousand and One Nights.'' In most cases, however, authentication occurs by biometric traits and also by some form of multi-factor authentication, even in a pre-technology world. For instance, when the hero is (or proves to be or reveals to be) the only one able to do something or to have something: 
\begin{itemize}
	\item When the disguised Odysseus returns home to Ithaca after 20 long years, his faithful dog Argos recognizes him by his smell and  his old wet-nurse Eurycleia by a scar he received during a boar hunt, and he finally proves his identity to his wife Penelope by being the only one able to string Odysseus' rigid bow and shoot an arrow through twelve axe shafts. Similar stories exist in Hindu mythology.
	\item Thor is the only one able to lift and wield the Mj\d{o}llnir hammer in Norse mythology (as well as in Marvel Comics and in Marvel Cinematic Universe).
	\item Arthur is the only one able to pull out the sword Excalibur from a stone, thereby proving to be the rightful king of Britain.
\end{itemize}
Similarly, mythologies and fairy tales also contain examples of \emph{masquerading attacks}, in which the attacker poses as an authorized user. In the Internet, this would occur by the attacker using stolen logon ids or passwords, in mythology it occurs by the attacker magically or divinely shape-shifting into somebody or something else (e.g., in the ``Epic of Gilgamesh'', in the ``Iliad'' or in Ovid's ``Metamorphoses''), whereas in fairy tales it often occurs via simpler means, e.g., the wolf pretends to be mother goat by eating honey to soften his voice and by smearing flour over his feet to turn them white in ``The Wolf and the Seven Young Goats'', a story published by the Brothers Grimm in the first edition of ``Kinder- und Hausm\"archen'' in 1812~\cite{Grimm}. Similar tales have been told in other parts of Europe and in the Middle East.


Finally, there are also examples of cases in which authentication fails not because the attacker is actively trying to deceive his victims, but rather because the victims (consciously or not) want to be deceived, as in the story of Martin Guerre, a French peasant who, in the 1540s, was at the center of a famous case of imposture: several years after Martin Guerre had left his wife and village, a man claiming to be him appeared in the village and lived with Guerre's wife and son for three years, before eventually being accused of the impersonation. The case of Martin Guerre has been popularized in literature~\cite{Lewis-Guerre,Sciascia-sentenza}, film~\cite{MartinGuerre,Sommersby}, musical~\cite{MartinGuerre-musical}, as well as in plays and operas, and has also been the subject of scholar investigations~\cite{Davis-Guerre}. Many of these works portray how, for different reasons, some people, including his wife, authenticated the stranger to be Martin Guerre even though they suspected him to be an impostor or even knew him to be one.
A similar case happened in Italy in the late 1920s: the Bruneri-Canella case concerned an amnesiac patient of the Mental Hospital of Collegno\footnote{Collegno is a small town in the North-West of Italy and the case is known in Italy more colloquially as the ``Smemorato di Collegno'', i.e., the amnesiac of Collegno.}, who was  identified first by the Canella family as the professor Giulio Canella, who had gone missing in action during World War I, and then by the Bruneri family as the fugitive petty criminal Mario Bruneri. After several inquiries and trials, the court found that he was indeed Bruneri, but the Canella family kept claiming he was Giulio Canella and he lived with Canella's wife, Giulia Canella, in exile in Brazil until his death in 1941. 
Also in this case there were people who wanted to (wrongly) authenticate the amnesiac: some newspapers stated that actually Giulia Canella herself was ultimately convinced that the amnesiac was not her husband, but she had to keep pretending otherwise to avoid a major scandal. And  this case too inspired literature~\cite{Sciascia-memoria}, plays~\cite{Cometumivuoi} and films~\cite{AsYouDesireMe,Unoscandaloperbene}, which can be used to explain authentication and related attacks.

Such ``popular'' explanations are not meant to replace the mathematical definitions and explanations, nor the facts and skills mentioned in the above quote of by Zazkis and Liljedahl. Although finding such popular artwork examples is challenging, the synergy of telling and showing via these examples, which laypersons will likely be already familiar with, can help go beyond the mere facts and skills by making them more intuitive, more accessible, more interesting, more rewarding.

\section{Conclusions}

Films have been used to explain and teach different disciplines such as 
philosophy~\cite{Ariemma17,Cabrera07,Mordacci17},
history~\cite{Marcus18},
social sciences~\cite{Russell12},
management and organizational behavior~\cite{Champoux00b,Champoux00a},
international relations and politics~\cite{EngertSpencer04,Valeriano13},
and mental health~\cite{Rubin12}. 
In addition to my own research, initial investigations have also been carried out for cybersecurity in~\cite{BlascoQuaglia18}, but more work is needed to explore the full potential of popular films and artworks for cybersecurity.

I have described four categories of such artworks and provided some examples for each of them. I have been collecting a database that includes many more examples and has benefited not only from the help of colleagues and friends (such as Diego Sempreboni, Sally Marlow and Gabriele Costa), but also from the lists in~\cite{Bond1946,Dooley2016} as well the entries in the Internet Movie Database IMDB. I am keen to include examples from less considered artworks such as plays and music, and to investigate the use of new artworks that have been created on purpose or that are created live during a presentation. The collaboration that I have initiated with artists and curators (such as Hannah Redler Hawes and Alistair Gentry) will be very fruitful to that end.


\section*{Acknowledgements}

This work was supported by the King's Together Multi and Interdisciplinary Research Scheme, King's College London, UK. Thanks to Giampaolo Bella, Gabriele Costa, Alistair Gentry, Sally Marlow, Hannah Redler Hawes and Diego Sempreboni for their invaluable contributions, and to Michele Emmer, Ashwin Mathew, Alessandra Di Pierro and Aldo and Claudia Vigan\`o for many useful suggestions.

%
%
%

\end{document}